\documentclass[sigconf,compress]{acmart}

\usepackage{booktabs} 

\setcopyright{rightsretained}
\acmConference[PriSC'19]{Principles of Secure Compilation}{January 13--19, 2019}{Lisbon, Portugal} 
\acmYear{2019}
\copyrightyear{2019}
\acmPrice{15.00}

\usepackage{url}
\usepackage{xcolor}
\usepackage{subfigure}
\usepackage{alltt}
\usepackage{fancyhdr}
\usepackage{amstext}
\usepackage{amsmath}  
\usepackage{amssymb}  
\usepackage{amsthm}
\usepackage{mathtools}
\usepackage{stmaryrd} 
\usepackage{newfloat} 
\usepackage{framed}   

\makeatletter
\@ifpackageloaded{txfonts}\@tempswafalse\@tempswatrue
\if@tempswa
  \DeclareFontFamily{U}{txsymbols}{}
  \DeclareFontFamily{U}{txAMSb}{}
  \DeclareSymbolFont{txsymbols}{OMS}{txsy}{m}{n}
  \SetSymbolFont{txsymbols}{bold}{OMS}{txsy}{bx}{n}
  \DeclareFontSubstitution{OMS}{txsy}{m}{n}
  \DeclareSymbolFont{txAMSb}{U}{txsyb}{m}{n}
  \SetSymbolFont{txAMSb}{bold}{U}{txsyb}{bx}{n}
  \DeclareFontSubstitution{U}{txsyb}{m}{n}
  \DeclareMathSymbol{\aleph}{\mathord}{txsymbols}{64}
  \DeclareMathSymbol{\beth}{\mathord}{txAMSb}{105}
  \DeclareMathSymbol{\gimel}{\mathord}{txAMSb}{106}
  \DeclareMathSymbol{\daleth}{\mathord}{txAMSb}{107}
\fi
\makeatother

\def\authorone{Peter T. Breuer}
\def\authortwo{Jonathan P. Bowen}

\def\E{{\cal{E}}}
\def\enc[#1]{\E[#1]}
\def\O{{\cal{O}}}
\def\D{{\cal{D}}}
\def\dec[#1]{\D[#1]}
\def\QED{\raisebox{0.8ex}{\framebox{\kern0.2ex}}}
\def\C#1{{\mathbb C}[#1]}

\DeclareFontShape{OT1}{cmtt}{bx}{n}
     {
      <5> <6> <7> <8> <9>
      <10> <10.95> <12> <14.4> <17.28> <20.74> <24.88> cmbtt10
      }{}
\DeclareFontShape{OT1}{cmtt}{b}{n}
  {<->sub * cmtt/bx/n}{}

\theoremstyle{plain}
\newtheorem*{theorem*}{\sc Theorem} 
\newtheorem*{lemma*}{\sc Lemma}
\theoremstyle{acmplain}

\makeatletter
\def\blfootnote{\xdef\@thefnmark{}\@footnotetext}
\makeatother

\makeatletter
\DeclareRobustCommand*\cal{\@fontswitch\relax\mathcal}
\makeatother

\DeclareFloatingEnvironment[fileext=box,placement={!tb},name=Box]{mybox}

\begin{document}
\pagestyle{headings}  

\title{(Un)Encrypted Computing and Indistinguishability Obfuscation}

\author{\authorone}
\affiliation{
          \institution{Hecusys LLC}
          \city{Atlanta}
          \state{GA}
          \country{USA}
}
\email{ptb@hecusys.com}
\author{\authortwo}
\affiliation{
        \institution{London South Bank University}
        \city{London}
        \country{UK}
}
\email{}

\renewcommand{\shortauthors}{\colorbox{black}{P. Breuer et al.}}
\begin{abstract}
This paper first describes an `obfuscating' compiler technology
developed for encrypted computing, then examines if the trivial case
without encryption produces much-sought indistinguishability
obfuscation.
\end{abstract}

\maketitle

\section{{Introduction}}
\label{s:Intro}

\noindent
{\em Encrypted computing} \cite{BB13a} means running on a processor that
`works profoundly encrypted' in user mode, taking encrypted inputs to
encrypted outputs via encrypted intermediate values in registers and
memory.  Encryption keys for such processors are installed at
manufacture, as with Smartcards \cite{SmartCard}, or uploaded in public
view to a \mbox{write-only} internal store via a Diffie-Hellman circuit
\cite{buer2006cmos}, and are not accessible to the operator and
operating system, who are the unprivileged user's potential adversaries
in this context.  Prototype processors supporting encrypted computing
include HEROIC \cite{heroic}, CryptoBlaze \cite{cryptoblaze18} and the
authors' own KPU (Krypto Processing Unit) \cite{BB18b}.  The latter,
clocked at 1\,GHz, measures on the industry-standard Dhrystones
benchmark \cite{Weicker84} with AES (American Encryption Standard)
128-bit encryption \cite{DR2002} as equivalent to a 433\,MHz classic
Pentium.  HEROIC and CryptoBlaze use the much slower Paillier
\cite{Pail99} (partially homomorphic\footnote{$\E[x{+}y]=\E[x]*\E[y]
\mod m$, Paillier encryption $\E$, public key\,$m$.}) encryption at 2048
bits or more.

A context in which an attack by the operator may be a risk, for example,
is where scenes from animation cinematography are being rendered in a
server farm.  On-site operators have an opportunity to pirate for profit
portions of the movie before release and may be tempted.  Another
possible risk scenario is processing in a specialised facility of
satellite photos of a foreign power's military installations to reveal
changes since a previous pass.  If an operator (or a hacked operating
system) can modify the data to show no change where there has been some,
then that is an option for espionage.  A {\em successful attack} by the
operator in both cases is one that discovers the plaintext of user data
or alters it to order.  It is shown in \cite{BB18c} that, given that the
encryption is independently secure:
\begin{theorem*}
There is no method, deterministic or stochastic, that can read a bit of
plaintext data from a program trace or rewrite the program to generate
a target bit to order, with any probability above random chance.
\end{theorem*}

\noindent
The method that is ruled out is not restricted to polynomial
complexities (in the encryption block size), but apart from that
and the extra hypothesis, that is classic `cryptographic semantic
security' \cite{Goldwasser1982} and the meaning here is `encrypted
computing does not compromise encryption'. The result also depends
on stochastically-based compilation, as follows.

An appropriate `obfuscating' compilation $\C{-}^r$ for encrypted
computing and ANSI C \cite{ansi99} takes an expression $e$ of the source
language and compiles it to machine code {\it mc} that puts it in
register $r$ but deliberately misses the nominal value by an offset
$\Delta e$ that it generates, as set out in \cite{BB17a}:
\begin{equation}
\C{e}^r = ({\it mc},\Delta e)
\end{equation}
The operational semantics of the generated object code {\it mc} is such
that it changes processor state $s_0$ to an $s_1$ with a ciphertext
$s_1(r)$ in $r$ whose plaintext value beneath the encryption $\E$
differs by $\Delta e$ from the nominal value\footnote{Syntax $e$ vs.\
value $s(e)$ are not distinguished for succinctness here.} $e$:
\begin{equation}
s_0 \mathop\rightsquigarrow\limits^{\it mc}  s_1 ~\text{where}~ s_1(r) = 
\E[e + \Delta e]
\end{equation}
The following lemma encapsulates the compiler specification:
\begin{lemma*}
Object codes {\it mc} from the same source are identical apart from
embedded (encrypted) constants.  The runtime traces are also identical
apart from the cipherspace values read and written, such that, for any
particular plaintext 32-bit value $x$, the probability across different
compilations that $\enc[x]$ is in a given register or memory location at
any given point in the trace is uniformly $1/2^{32}$, independently to
the maximum extent permitted by copy instructions and loops in the code.
\end{lemma*}

\noindent
The proviso is because a plain copy (`mov') instruction always has
precisely the same input and output, and a loop means the `delta'
variations introduced by the compiler must be the same at the beginning
as at the end of the loop.

The set of deltas generated by the compiler as above, one per register
and memory location per point in the control graph is an {\em
obfuscation scheme} $\O$.  The user knows the scheme, so can interpret
the program output after decryption, but `the processor' does not know
it, nor does the operator or operating system, in addition to not having
access to the encryption key.  The compiler can also generate schemes in
which the input and final output deltas are zero.  Then different object
codes that look the same apart from constants, whose traces look the
same apart from values written and read, end up with the same (correct,
encrypted) values from the same inputs.  That is a kind of obfuscation,
limited to different programs from the same source by the same compiler.

That is the established theory.  The question here is [i] if encryption
$\E$ is really necessary, and [ii], what happens if a virtual machine
(VM) is compiled to machine codes $v_1'$, $v_2'$ conforming to
obfuscation schemes $\O_1$, $\O_2$ respectively along with data
$\E[p_1]$, $\E[p_2]$ interpreted by $v_1'$, $v_2'$ as programs, and data
$\E[d]$ for them.  The data $d$ is the same for both because it is
compiled with delta zero, but deltas for $p_1$, $p_2$ differ.

Looking at [ii] first, by definition, on a processor for encrypted
computing:
\begin{equation}
v_1'(\E[p_1],\E[d]) = \E[o],  \quad v_2'(\E[p_2],\E[d]) = \E[o]
\end{equation}
Examining that for [i], on an unencrypted platform, where $v_1$ and
$v_2$ are the versions of the programs that embed unencrypted
constants in the machine code instructions instead of encrypted
constants, gives
\begin{equation}
v_1(p_1,d) = o = v_2(p_2,d)
\end{equation}
Both programs produce the same results.  The traces are those of the
repetitive virtual machine cycle interpreting the incoming data as
instructions and modifying components of its state (which may be thought
of as a finite array representing memory) to suit.  They take the same
time to finish.  The different obfuscation schemes will have scrambled
the access patterns (the compiler changes the delta in force for a
pointer or array index at every increment to it).  It is impossible to
tell which is which, but both come from the same source.

What happens with different source codes with the same functionality?
One expects traces to be different lengths and show different patterns,
but suppose now that instructions are ordered/numbered arbitrarily in
$p_1$, $p_2$ and there is only one instruction type for the VM: add,
compare and jump:
\begin{equation}
\label{e:ins}
L_0: {\bf if}~(Y = X + A) < Z + B~{\bf goto}~L_1~{\bf else}~{\bf goto}~L_2
\end{equation}
There, $A$, $B$ are constants, $X$, $Y$, $Z$ are program variables,
$L_0$, $L_1$, $L_2$ are instruction locations.  Comparisons and
additions wrap in the 2s complement arithmetic.  This one instruction
suffices for all computation: it is the instruction in HEROIC's
encrypted `one instruction computing' (OIC) system.

An instruction \eqref{e:ins} can also be interpreted via $X{=}x{+}a$,
$Y{=}y{+}b$, $Z{=}z{+}c$, where $x$, $y$, $z$ are `virtual variables'
and $a$, $b$, $c$ are secret constants, as follows:
\begin{equation}
\label{e:ins2}
L_0: {\bf if}~(y = x + A') < z + B'~{\bf goto}~L_1~{\bf else}~{\bf goto}~L_2
\end{equation}
where $A'=A+a-b$, $B'=B+c-b$.

Our compiler for encrypted computing already randomly generates deltas
$a$, $b$, $c$ by varying $A$, $B$, $C$ per instruction and manages the
deltas over the course of translation of a given program to maintain the
intended source code semantics.
Here the `virtual variables' $x$, $y$, $z$ in \eqref{e:ins2} have no
physical existence so cannot be directly observed, but that is
abstractly the same situation as on an encrypted computing platform,
where encryption protects register and memory contents.  Theory
developed for the compiler in the encrypted computing context then
applies.
The Lemma affirms the compiler can design instruction constants $A$,
$B$, $C$ to vary secret $a$, $b$, $c$ deltas independently and
arbitrarily.

Only the user who knows those `program keys' $a$, $b$, $c$ can interpret
the program's functioning, and they can only run the program code
correctly as data $p$ for a virtual machine $v$ that is supplied with it
if they know the `master key' consisting of the obfuscation scheme $\O$
for $v$.  The VM $v$ can have been customised for $p$ with a different
interpretation unit for each instruction internal to $p$, plus the
compiler's obfuscations.

\section*{Summary}
A compiler for encrypted computing, on each recompilation of the same
source, generates object code of the same structure for which the
runtime traces also have the same structure but for which the data
(beneath the encryption if there is any) differs at each point in the
trace and memory with flat stochastic distribution, independently to the
maximal extent.

The full paper looks at what that compiler does in an unencrypted
context where it translates source to machine code for a virtual
machine.  Instead of encrypted registers and memory, virtual content is
hidden via secret deltas from physical values.  Translation of a
particular source code constructs are examined very carefully to see if
this mechanism can possibly amount formally to `indistinguishability
obfuscation'.

\renewcommand{\baselinestretch}{0.93}
\bibliographystyle{ACM-Reference-Format}
\begin{small}
\bibliography{prisc2019}
\end{small}
\renewcommand{\baselinestretch}{1}

\end{document}